Name:       José Carlos Bermejo-Barrera

Address:    Universidad de Santiago de Compostela
            Facultad de Geografía e Historia
            15782 Santiago de Compostela
            Spain

Telephone:  011 34 981 563 100 Ext.12562

Fax:        011 34 981 559 941

E-mail:     jcbermej@usc.es

Words:      8,600

A NARRATION IS NOT AN EQUATION:

METAPHYSICAL PRINCIPALS OF STANDARD COSMOLOGY

We live in a world where science is considered to be the only valid form of knowledge; furthermore, it is taken for granted that such a thing called science actually exists. Science can be characterized by having a method that defines it and establishes the limits between scientific knowledge and other forms of knowledge that do not posses this universal validity -- which is the most specific characteristic of scientific knowledge. This science, characterized by the mastering of a method (a method which, interestingly enough, philosophers of science are ever less capable of defining), possesses yet another defining characteristic: unitary knowledge. The unity of science, an old aspiration of the Vienna Circle, can be defined by two assumptions. The first of these would be of a logical nature, and can be formulated in the following manner:

1) There is a body of knowledge that can be denominated by the noun *science*. If these types of knowledge form part of a set, they must posses at least one attribute in common. This attribute can be either: (a) Formal, where all knowledge is configured in the same form or by the same method; or (b) Material, where all knowledge is part of a common subject, and that subject is the universe, or field, to which that knowledge refers.

We can categorize the second assumption in the following manner because it is shared -- without being clearly formulated -- by all scientists:

2) Science is an ordered conjunct, within which its members are not only perfectly ordered, but also hierarchized in such a way that science can account for the totality of the universe because it is capable of developing explanations in a hierarchical manner, beginning with the most simple and ending with the most complex.

This hierarchy of scientific knowledge -- which incorporates harmoniously what was known as the *Great chain of being* within the Aristotelian and Scholastic traditions and covered everything from unformed and inert matter to God -- is now presented to a public composed of those of us who are not scientists as well as those who are. Among the latter are some who aspire to offer a global synthesis of scientific knowledge in the format of a big history, or a big narration, which begins at the supposed moment of the origin of the cosmos and ends at the present time, thus giving an account of the reason of being of the universe as well as of humankind.

This historicization of the universe, which would permit the integration of all knowledge, from topology and geometry to sociology and history, passing through physics, astronomy, chemistry and biology, is becoming more and more accepted as a perfectly coherent system of knowledge, and, like all kinds of overly coherent knowledge, attempts to usurp reality. These types of systematizations are currently offered in various sorts of books, some written by popularizers, others written by eminent scientists who proceed from fields such as physics, chemistry, biology, or even history or sociology. All these books start off from a common scheme, narrating a "big history" from the Big Bang until today, as Fred Spier's title indicates, a book that we can consider a Vulgate within these types of narrative creations.[1]

If we wish to search for the origins of these big histories we can resort to two types of sources, which interestingly enough converge in a common root. On the one hand, we find books written by historians who not long ago understood that human history cannot be dissociated from the history of the Earth, and also that this planet is no more than a miniscule part of the cosmos. Such is the case of H. G. Wells, who wrote the first book which focused on human history within the framework of an ecologic context, attempting to find large global tendencies; this book has been enlarged and republished numerous times.[2] Another such author is William H. McNeill, who carried out a similar endeavor, although not from the perspective of a novelist who was fond of history, as was the case with H. G. Wells, but from the perspective of a professional historian.[3]

Wells and McNeill completed their work long before the arrival of the term "standard cosmology." By integrating relativity with quantum mechanics, this cosmology would later be considered definitive. Once this type of cosmology was systemized, other historians such as David Christian went on to integrate these two narrative structures believing that they were achieving the integration of cosmology and history, which is really nothing more than what was previously known as natural science or science of culture or history.[4]

What we will attempt to accomplish hereafter is an analysis of the assumptions shared by all the authors of these types of accounts that -- following Fred Spier's example -- we shall call "big history." These assumptions are strictly of a metaphysical nature and therefore tell us more about the writers than about the contents of the narratives.

Our work can be placed along the same line of thought initiated by John Dupré, starting out from the principle according to which the idea that the universe is something ordered does not refer to the universe itself; instead, it refers to an anthropological assumption that is indispensable for defending the idea of the unity of knowledge -- which was once theological and philosophical -- and of the unity of science. Dupré prefers to talk about the "disorder of things," a slogan that we should only accept rhetorically because, in reality, neither order nor disorder are attributes of the universe itself. If Dupré coined this expression for the title of his book, he did it to a certain extent to explain he was against supposedly global explanations of the universe, which posses a series of common characteristics, both within those models named deterministic as well as within other models named probabilistic.[5]

It is well known that the person who most clearly symbolizes the determinist ideal in cosmology is Pierre Simon Laplace. In the early nineteenth century, with his "system of the world," Laplace attempted to give credibility to the idea that the universe is perfectly ordered and coherent, and that if we were to understand all the elements that composed it, we would be able to regress to each and every past moment and also predict the future. The universe would thus be the gigantic mechanical clock the European Enlightenment was fond of imagining. However, upon reading Laplace we find that his determinism limits itself to the solar system, and that it conceals numerous problems. There are facts like Mercury's perihelion that cannot be explained; furthermore, Laplace's mathematical model becomes useless when more than two planets enter into the calculations, thus raising the problem of the "three bodies." Nevertheless, Laplace maintained his determinist and cosmologic faith, which is basically shared by present day cosmologists, biologists, and historians. We are not going to attempt to explain the problems of the "three bodies" of the current cosmologies, but simply shed light on the obscure points where lack of evidence is covered by the appeal to a series of metaphysical principles, of which scientists and historians are not totally cognizant.[6]

All contemporary cosmology is based on what is known as the *Anthropic Principle*, according to which it should be possible to explain how humans have come into being and are able to exist in relative stability within the framework of the universe. The anthropic principle is nothing more than a reformulation of the old Scholastic principle, which later became a Rationalist principle known as the *Principle of Sufficient Reason*, recently analyzed and reassessed by Alexander R. Pruss.[7] According to this principle: *nihil est sine ratione*, in other words, we can account for everything that exists. As Arthur Schopenhauer mentioned in his doctoral dissertation, *On the Fourfold Root of the Principle of Sufficient Reason*, this principle can be confused with the principle of causality, passionately defended by all the champions of unified science and of the rationality of the universe, as is the case of one of the best known, the Argentine-Canadian philosopher Mario Bunge.[8]

In the analysis we will attempt to develop, we start off from precisely the opposite premise: *omnis est sine ratione*, that is, the reason of things does not reside within the disorder of things themselves, but within the language. Reason and order are not attributes of the universe, but of thought, and the only way to be coherent, if we want to defend the anthropic principle and the principle of sufficient reason, is to assume that God exists and is a thinking being who created the universe passing onto it His best attribute: the attribute of rationality.

In our analysis, which can be labeled a "reassertion of contingency," we start off from an epistemological principle established by the physicist Pierre Duhem in 1914, and reassessed recently by Nicolas Rescher, which states that the amplitude and the extension of a scientific theory are inversely proportional.[9] According to this principle, a theory of everything, a final theory, like those that many physicists are fond of -- including Stephen Hawking and certain "superstring" theorists such as Brian Greene -- are characterized for being wholly inaccurate.[10]

We shall attempt to describe the obscure points of "standard cosmology" by following its own narrative thread. Standard cosmology is a narration, constructed on the basis of various types of scientific knowledge. As is the case with all narrations, we can distinguish the following parts: (1) Every narration takes place on a stage -- in this case the stage is the universe. (2) Every narration has a protagonist -- in this case the protagonist is not the universe, it is humankind. (3) The protagonist needs *time* to develop an action -- here time is identified with history itself, which Hawking actually calls "history of time," but time is not the protagonist of the story, simply a condition for the story to occur. (4) The protagonist develops his or her action with the aid of certain means -- in this case those means are the different structures of matter, which follow the road of ontological hierarchy. (5) Every narration concludes with an *end* -- in this case the end is the present time, although some narrators such as Stephen Hawking and Paul Davies are fond of constructing models of the destruction of the universe, or they attempt to revive the myth of the eternal return with universes in constant expansion and contraction.[11] Some even imagine bizarre contradictions in terms such as "multiverses," which are the totalities that compose the whole -- something which cannot be, and, what's more, is impossible, as a famous bullfighter once said, a statement with which David Deutsch would not agree.[12]

In our narration we shall follow a number of different writers who sum up the "standard cosmology," authors such as Steven Weinberg, Harry L. Shipman, Roger Penrose, Robert Geroch, Michael Disney, Harald Fritzsch, and even Stephen Hawking himself.[13] The fact that the authors have omitted the mathematical apparatus in these books, and the fact that neither I nor many other readers would be able to understand the mathematics had it been included, does not interfere with our ability to understand the theory, for two different reasons. First of all, because the authors themselves state that they can explain the essential aspects without resorting to mathematics, and secondly because we are interested in the metaphysical assumptions and these cannot be mathematically formulated. If they could be, they would not be metaphysical. And if the essential aspects of the theory could not be explained in ordinary language, then it would fall within the field of the ineffable and form part of those experiences that cannot be shared linguistically.

To understand this story we must begin with the modest observation made by astronomer Edwin Hubble many years ago. Hubble, who was not a theorist but basically an observer, reported that the spectrum of light of all the galaxies observable from the earth was shifting toward red, which is to say that the band of the spectrum corresponding to red was becoming wider. As it is a well known fact that a "redshift" occurs when a light-emitting object moves away from an observer, Hubble concluded that all the galaxies where moving away from us. If they were moving away, this meant that they were separating, and if they were separating, it meant that at one time they had been closer, thus the universe had to be expanding. Our reference books explain -- so that we can understand the concept -- that the universe is like a sponge cake with raisins which has been placed in an oven, and as it begins to increase in size, the raisins, which would be our galaxies, move away from one another.

If we go back in time clustering galaxies, we must arrive at a point when all of them were together. We can formulate the "primeval scene" of our cosmic history by starting off from this idea and uniting the theory of relativity with the theory of quantum mechanics. Einstein established that the universe could be understood as an enormous field of forces, in which the fundamental ones were -- at his time at least -- gravity and electromagnetism. The universe would thus be a structure formed by space and time, and the configuration of that space-time would have been the result of the density of matter. If we construct a mathematical model basing ourselves on this theory and on Hubble's observation, we can conclude the following: According with the field equations of the so-called general relativity, a situation might arise in which space-time could collapse, with space disappearing in a point without dimensions.

Within these field equations the so-called *singularities* are produced, which are cases when these equations, so to speak, collapse. Within the singularities of relativity, space loses its volume; consequently, matter, which continues to exist, reaches an infinite density because density is equal to matter divided by volume. These singularities, predicted by Einstein, are now called *black holes* -- physical phenomena that, for our narrators, are as prestigious as they are exotic. The initial phase of the universe can, therefore, be considered an enormous black hole which expanded until reaching the present state. The fact that Einstein predicted the existence of black holes constitutes a great part of their prestige.[14] What Einstein could not foresee, however, was that all this could be integrated with quantum mechanics, which he always disliked for aesthetic reasons because it disrupted his idea of a harmonious universe.[15]

Thanks to the development of quantum chromodynamics (the physical-mathematical theory of quarks) and later the development of the theory of superstrings, quantum mechanics was able to integrate the relativist model of the universe with the history of elemental particles in the following manner: A concentrated universe of almost infinite density is a very hot universe in which matter is under great pressure. Under those conditions of temperature and pressure only certain types of particles can exist. Quarks had their origin in the superstrings which, of course, are hypothetical, since, by definition, they cannot be observed -- as is the case of the quarks, which are "confined" within particles. Through the combination of quarks: Top, Bottom, Up, Down, Strange and Charm, heavy particles called baryons would have been formed, and thus, as the universe continued to cool and expand, what is possibly the most important event in the history of the cosmos took place, the appearance of hydrogen -- the first type of atom.

The combustion of hydrogen and its transformation into helium, according to the equation: $E=mc^2$, which results from the contraction of the gas due to gravitational force, would have produced the stars and the galaxies, which continued to form part of the simultaneous processes of expansion and cooling down. From the cooling down of the fragments of a star -- which in our case would be the sun -- the planets would have been formed, including the Earth. After the cooling down process had brought about new elements of the periodic system in the Earth, the first molecules appeared and these would have made life possible. Life had its own dynamics, culminating in the arrival of humans. The engraving on the final page, taken from Harald Fritzsch's *Quarks: The Stuff of Matter*, provides a straight-forward display of this narrative scheme.

We will finish narrating the story before pointing out its problems. Within the world of chemistry, the distinction between organic and inorganic compounds was regarded almost as a dogma. This dogma was demolished with the synthesis of urea. The jump from inorganic to organic was only a first step. Starting out with organic compounds, scientists later discovered the large molecules -- studied by biochemistry -- which make up all living beings.

It is not known how or where the synthesis of molecules that can copy each other, or duplicate themselves, first took place. These molecules are the genes; they synthesize proteins, and living beings are constituted from them. Genes are composed of four bases (adenine, cytosine, guanine, and thymine), which can recombine in almost infinite ways, thus accounting for the diversity of life forms. Genes are formed basically from two types of molecules: RNA and DNA, with DNA being fundamental for the development of life as we know it.

It is not easy to explain the origin of life. According to Fred Hoyle and N. C. Wickramasinghe, the physical conditions for such synthesis could only have occurred outside the Earth, with the necessary molecules arriving on a meteorite.[16] Other authors, including Paul Davis, Jesús Mosterín, and Roger Penrose, do not fully accept the contingent act of the arrival of one or more meteorites.[17] Nevertheless, the one thing all of them agree upon is that the advent of life had to be a contingent event, just as all species of life are contingent, merely developing some of the possibilities of an almost infinite number of genetic programs.

Continuing with our story we can find that the history of life went from unicellular to pluricellular organisms, and from asexual to sexual reproduction, in progressively increasing complexity culminating with human beings as the superior life form. Here we would have to accept the fact that standard cosmology is manifested by the theory of evolution, a theory which is more or less plausible; however, due to religious or ideological reasons it is often cited not as a theory, but as a fact. Stephen Jay Gould has done an extensive study of its history and structure.[18] Basing ourselves on his work, and on the work of Ernst Mayr, we will now put forward some of the problems presented by this theory, showing the weakness of the principle of sufficient reason, as opposed to the power of contingency.[19]

Mayr points out that within biology there are three paradigms -- scientific systems -- that function in parallel and are sometimes difficult to merge together. There is a descriptive biology whose mission is to catalogue the millions of life forms, the exact number of which is unknown, although it is estimated to be approximately three million at present. This system is based on morphology, and some surprisingly simple and beautiful results can be obtained -- such as those of Sir D'Arcy Wentworth Thompson -- observing the underlying geometric proportions of many life forms, which seem to follow mathematical laws; nevertheless, morphology is the realm of plurality, contingency, and continuous variation.[20]

Morphology and genetics are most probably related, but it is impossible to integrate them at present, and perhaps always will be, due to the impossibility of explaining billions of possible forms and combinations. The morphologic and genetic debate needs to be linked to the evolutional one in the same manner, but it is not so simple to make them coincide. Gould has pointed out that Darwin established the almost uselessness of the fossil registry for verification of his theory. According to Darwin, this was due to the fact that the transmission of fossils is arbitrary and dependent on totally contingent geologic and chemical circumstances; thus, fossils could neither refute nor endorse a theory that was essentially built upon suppositions arrived at from morphologic observations.

Gould points out this contradiction between genetic and evolutionary logic. As opposed to the geneticists who are fond of speaking of selfish genes and genetic logic as the key to evolutionary dynamics, this author upholds the old notion of the organism. The organism possesses a structure and logic that in many cases subordinates the genes. Evolution would then be the evolution of organisms and species, and not primarily of the genes. The organism belongs to the realm of morphology, and morphology to the realm of contingent variety; thus, in this case as well, the omnipotence of the "reason of history" would diminish -- to use a Hegelian term.

Gould, furthermore, proposes a much more contingent model of evolution as opposed to a determinist model marked by necessity. He is the author of a theory called "punctuated equilibrium," according to which, organisms which adapt themselves perfectly to their environment become fragile and are not able to respond to catastrophic contingencies. When these occur, the less adapted, or marginalized, organisms are the ones who survive, adapting themselves to the new environment until the time when they are displaced by other marginalized organisms. This would have been the case of the extinction of the dinosaurs 65 million years ago and the success of our mammal ancestors.

As opposed to a rather dogmatic evolutionism, which, long before Darwin published his *Origin of Species*, was already a reflection of numerous ideas of Victorian political economics endorsing the survival of the fittest,[21] Gould emphasizes a non-providential evolutionism, not based on the principle of sufficient reason, but conceived much more contingently within the scope of classical historiography, or within that of European historicism.

The problems of biological knowledge are much more complex than those of physical and chemical knowledge. There are open questions, such as the interaction between organic and inorganic substances, which seem to form part of the same system and make the theory of evolution and the concept of biology itself much more complex. James Lovelock has raised some of these questions. His observations, leaving aside the exaggerations, seem to hold much truth in them.[22]

If we move on from the field of evolution to that of human history, the matter becomes even more complex, because we would be required to touch upon the fields of philosophy and theory of history, which are not now our objective. Nowadays, it seems quite clear that the former idea of progress might have its lights and shadows, and the belief that we are at the highest phase of historical development -- just because it is the latest -- does not make much sense, precisely because the integration of human history within the ecologic field does not stop the exposure of the fragility of our supposedly rational and apparently secure world.

Having provided a broad outline of our cosmologic history, we shall now attempt to clarify the manner in which culture talks through science, as Pierre Thuillier stated, and in this case we could refer to an implicit lack of philosophical culture.[23] Let us return philosophically to the beginnings. Standard cosmology is presented to us as a very coherent narration in which a succession of perfectly ordered events takes place across a long time span, from the beginnings of the universe until the present. Within this succession of events we ascend on a scale where the simple evolves into the complex, following steps very similar to those of the old chain of being.

This succession of events is presented as the result of multiple scientific investigations, which converge among each other, to the point where we are told that we are faced with theories of everything that explain all kinds of facts, from quantum data to the human mind, as occurs in the book by Roger Penrose.[24] The matter, however, is a bit more complex, since these theories are only capable of explaining a small part of this long chain of events. Furthermore, the various scientific theories -- which apparently form part of a whole -- not only do not converge, but in many cases are incompatible to each other, as with the three simultaneous paradigms of biology previously mentioned.

Standard cosmology is a narration constructed on the basis of a series of scientific theories that attempt to group together facts and events of a very diverse nature. One cannot affirm that the narration is true for being based on scientific knowledge. No narration is true or false, only more or less credible, its credibility depending on the cultural or religious values of its age, and the knowledge the narration is capable of integrating within those cultural value systems.

The structure of standard cosmology is fully narrative, since it fulfills all the characteristics of a story we have enumerated earlier. In other words, it consists of an action, a stage, one or various agents, and the means necessary to achieve an end. The ending of our cosmologic narrative is the present time. This type of finale, contrary to what occurs in other providential stories, does not assert that we have arrived at the closing of history from where we can no longer advance, as was the case, for example, with nineteenth-century evolutionism.

We now possess enough ecologic and cosmologic knowledge to know that humankind is not necessarily the heir of the universe. On the contrary, our species seems to be terribly fragile and could face extinction from ecologic or cosmologic catastrophes, as occurred with the dinosaurs due to the impact of a meteorite. It is even possible to calculate the characteristics of a meteorite that could bring about our extinction.[25] Such successive extinctions are precisely the key, as we have seen, of Gould's theory of punctuated equilibrium.

Although it is true that on an ontological level we no longer share the faith in the perfection of the cosmos that would constitute the guarantee of our survival, that faith is maintained on the epistemological level. In other words, we believe that although we don't know everything (since science must progress for scientific work to make sense), we do know the essentials. Our bodies might very well be settled upon an unstable physical world, but our minds, on the contrary, advance through what Kant called "the sure path of science," a path from which we must not digress because it is capable of offering us sufficient explanations, presupposed by the anthropic principle, and by our old metaphysical principle of sufficient reason.

For the purpose of demonstrating that our standard cosmology is a narration, it would be interesting -- as with any other narration -- to clarify two flaws which it is based on, and which are usually hidden. These weaknesses may be of a linguistic or factual character. We will analyze both simultaneously.

Our cosmologists appeal to the Principle of Sufficient Reason (PSR), and here they come up against numerous problems. In the first place, they establish an initial stage for their story. That initial stage cannot be observed for two reasons; first, because it would have taken place some 15,000 million years ago; second, because obviously there could not have been any observers present.

If there is one thing we can be sure of ever since Einstein, it is that there is no absolute space or time, and that all physical observation depends on the reference system of the observer in the space-time continuum. This affirmation presupposes that all observations are contingent. If, furthermore, we consider that according to an old principle of quantum mechanics the observer modifies the object observed due to the interference of his observational devices -- which really means these devices construct the object -- we would, therefore, have to conclude that the Big Bang cannot be considered in any way whatsoever to be more than mere plausible fiction, and, moreover, a narrative type of fiction, and not scientific as it is presented to us.

The Big Bang -- the original event -- cannot be considered an event for reasons that even cosmologists point out. We are aware of only a small part of the matter that makes up the universe, and it seems that this matter is often confused with mass. It is said that a large part of the universe is made up of dark matter.[26] That matter cannot be observed because it does not produce light, or any type of radiation; waves -- of whatever type they may be -- are the only means by which information regarding the universe can be transmitted. If a large part of the universe is composed of unknown matter, we can say that the probability of the truth of the theory of standard cosmology would be equivalent to the ratio between known matter and dark matter: P(V) C.= known matter / dark matter. If this matter is supposed to make up the major part of our universe, then the probability of the truth of the theory of standard cosmology would be so low that no statistician would accept it as such (if, for example, it was less than 25%).

There are other problems as well. Cosmologists say there are four fundamental forces in the universe: gravitational, electromagnetic, strong interaction and weak interaction. All those forces are transmitted by particles, and it is possible to unite those forces to one another mathematically. The first accomplishment was the integration of gravity and electromagnetism. Three of those forces have been integrated, but it has not yet been possible to integrate all four, which is what the theory of the superstrings seeks to do. Even if they were to be integrated mathematically, however, the result would not necessarily be considered an undisputable fact. The particles that transmit gravity -- gravitones -- are a theoretical assumption, but they are practically unobservable. The same occurs with the results of the theory of the superstrings. If quarks are not observable, but only hypothetically deduced from mathematical models that attempt to explain certain experimental results, superstrings -- by definition -- are totally unobservable. There can be no observer nor can there be a spatial-temporal reference system, nor is it possible to construct an accelerator to test the viability of the theory; Brian Greene has stated that such an experiment would require an accelerator the size of the Milky Way.[27] This, besides being a technological *witticism* -- where would we get the material? -- is also an epistemological *witticism* -- where would the observer be situated? Let no one tell us that this is just an example, and that if we understood the complex mathematics of the superstrings we would be able to understand this metaphor *ad usum delphini* (to be understood even by children), because this statement has more serious consequences.

It is clear that the Big Bang cannot be considered a physical event, even though it is said to be observable because radiotelescopes can capture the "background radiation" that supposedly took place at the beginning of the universe. That is mere conjecture, not a fact. The characteristics of that radiation depend on the physical-mathematical predictions of a theory that only takes into account a minimal part of the possibilities. How can we be sure of the characteristics of the "background radiation" if we cannot observe the dark matter? Why do we say we know all types of radiation, when we only know a few, especially when we admit the possibility of the existence of matter that does not emit radiation. The difficulties posed by the facts are overcome by the apparent mathematical coherence and beauty of the theories involved. It is as if our cosmologists were to say that their theories are so beautiful that it is not worth risking a hard meeting with reality, which they themselves present. To understand this fact, which in no way should be attributed to lack of intelligence, nor to ill will on the part of our cosmologists, we should analyze their idea of mathematics.

Physicists say that they explain their theories linguistically because we -- the lay public -- would be unable to understand mathematical explanations. And that is certainly true. Hence, we, the lay public, can put forth the following question: Can ordinary language express in some way the contents of mathematics? If the mathematical contents are not merely formal (as in the case of standard cosmology), is there a place somewhere between mathematical language and ordinary language that can shelter cosmologic ideas?

We believe so, for the following reasons: Equations are the basic language of mathematics. These are structured on the basic concept of equality (=) and make a relationship among a series of dependent or independent variables. Those variables correspond to concepts: mass, energy, spin..., which must be unequivocally defined. In this case, those concepts, besides being formal constructs -- as is the case of algebraic concepts -- are assumed to refer to one or more facts. There could even be a case where a fact might be explained on the basis of various variables (as, for example, in an elementary equation: $F=m.a.$).

Mathematical syntax is perfectly constructed, thus it is said that science is a well-made language. In this language each mathematical proposition says exactly what it says, and there can be no room for metaphors or analogies -- two basic instruments of ordinary language. In the case of our cosmologists, however, we find that they jump over the limits of mathematical language, hiding facts that may contradict their stories, and bringing forth other facts that are most convenient. They do this with the purpose of constructing a narration in ordinary language, a narration, however, that they want to put forth as a very complex formal mathematical system, which would furthermore be substantiated by an enormous mass of data of all kinds: astronomic, physical, chemical, biologic,... Let's give a simple example. The cosmos and the universe are not scientific facts or mathematical concepts. There are no equations of the universe or theories of everything. The field equations of general relativity are mathematical constructs that draw up a structure of space-time in which it is assumed that the four known forces of physics will someday be integrated. The universe is not the same thing as the field equations.

The universe is the collection of everything that exists, or the collection of every kind of phenomena. The universe cannot be mathematically formalized, or turned into a fact or a field. This is so, especially if we take two things into account: first, the theory does not explain all known or possible facts; second, that same theory depends on the degree of development achieved by the different mathematical sciences, which must be considered provisional, and not definitive, unless we truly assume the SAP -- the strong anthropic principle.

An equation is a fundamental instrument, but not a fact. It is an instrument that lets physicists place a fact within a system and make sense of it. By means of an equation we can predict an event such as an eclipse.[28] This capacity amazed seventeenth-century Europeans and helped create the myth of the determinist and perfectly rational universe Laplace believed in. Yet physicists themselves confess that such a universe was nothing more than mere fiction created by the faith of the scientists of the Enlightenment. Physicists say that we live in a non-determinist world, in which the laws of statistics must replace the ancient laws of mechanics. Nevertheless, what is proposed on many levels is later forgotten when speaking of the history of the universe, offering a story supposedly substantiated by the great weight and authority of science, and endorsed by the solidity of mathematics and a large amount of empirical evidence.

While it is true that the universe is not a variable that can be included in an equation, or something that can be explained by a set number of equations, it is also true that our cosmologists base themselves on a series of assumptions that are mere metaphysics, offering them as the result of extremely complex scientific investigations. We previously explained that a singularity within a field equation is a mathematical fact according to which the equation would be of no more use, or to put it in lay terms, would collapse. A singularity is not a fact, even if the singularities of the field equations can be made to coalesce with black holes. A singularity is a frontier that establishes the limits where certain mathematical propositions are valid.

Standard cosmology's great singularity is the Big Bang, offered to us as a fact; however, it cannot be considered an observable event. The Big Bang is the limit between the language of mathematics and the language of cosmology. We are not saying that the Big Bang can be considered either as proof of the existence of God, nor as proof of the creation of the universe. Even such prominent physicists as Hawking have reiterated this. What we are saying is that the Big Bang is not a cosmological limit, it is an epistemological limit. The Big Bang is the limit of scientific languages. It is the collapse of these languages that now cannot continue to speak. There is nothing before or after the Big Bang, either tactically or conceptually. Nothing can be said about Big Bang scientifically, which would be the only way to speak about it, nor is there any other way of making any statements. That which can be said about all phenomena can only be said scientifically, as Ludwig Wittgenstein pointed out. The problem here is that what we are interested in talking about cannot be talked about. Wittgenstein called this the ineffable, because he maintained a private battle against philosophy and metaphysics.

We want to reassert the role of metaphysics as a way of speaking about these ideas, such as the idea of the universe. This is the reason why we are carrying out this analysis, where it will be necessary to continue observing the metaphysical assumptions of those who maintain there are no assumptions.

Standard cosmology not only confuses the field equations with the universe, but also presents as unquestionable the idea that all complex things may be reduced to simplicity. It is a dogma of physics that simple explanations are superior to complex ones. The Copernican system was better that the Ptolemaic system precisely because it was simpler. Physicists appreciate the value of simplicity, elegance, and harmony. Einstein said that he had always been convinced of the value of his theory -- even if it could not be proven -- because it was the most beautiful.[29] That is the reason why he did not like quantum mechanics and said that God did not play dice. Einstein's aesthetic ideals regarding physics turned out to be harmful to the development of the investigation, and cosmologists now criticize what they call the "great error" that undermined his investigations during the latter part of his life.[30]

Nevertheless, the aesthetic ideas regarding the value of simplicity and also the search for harmony continue to be present within standard cosmology. First of all, because it is the universe itself that becomes harmonious, not because it constitutes a perfectly stable system -- and this is recognized -- but because the "system of science" can account for the totality of the universe. At one time, human beings could feel secure in the universe because there were gods who guaranteed its order. Later, the laws of mechanics constituted the guarantee of our security. Now, it is science, understood as an anonymous and enormously complex system where the combination of technology and investigation, and the integration of technique, experimentation, and formal developments (mathematics and information systems) offer us a world governed by the definite system of truth, which would be the guarantee of all certainty and also of our happiness.[31]

The problem is that complexity takes revenge on simplicity, and experimental facts continually overwhelm the framework of theories and actual formal structures, which also end up collapsing. If the mere physical universe poses the problems mentioned earlier, the matter becomes more complex if we then bring up the old problem of the ontological spheres, which can now be reformulated as the problem of the limits of scientific languages.

If chemistry and physics exist, it is because chemical phenomena cannot be reduced to the physical sphere, even though there is a science called chemical physics (also known as "theoretical chemistry") which continues seeking this convergence, a convergence that is impossible because it seems likely that we will never be able to confine the almost infinite diversity of chemical facts and compounds within the same mold. If that ever occurred, it would be the end of chemistry and physics as we know them, and the beginning of another science -- or various other sciences -- that we cannot even begin to imagine. The language of chemistry starts where the language of physics stops, and if one cannot be reduced to the other, it means that the harmonious construct we call *science* presents us with numerous problems in its definition, a definition that is impossible to establish according to the assertions of certain "philosophers of science."

If we continue ascending the chain of being and move on from chemistry to biology, we would find ourselves in the exact same situation. The vital phenomena are not possible without a chemical and physical base -- life can only develop within a narrow margin of temperatures. Biology, however, cannot just be reduced to the language of chemistry, although biochemistry -- and genetics above it -- may try, of course, to do so. Chemical physics, biochemistry, and genetics are necessary endeavors and mean embarking upon a quest that is unattainable by definition. This does not, however, diminish its value. Karl Popper stated that science was an "unended quest" and that it was often more important to search for the truth than to find it. Popper was a philosopher with a scientific education. Modern scientists, on the contrary, have no philosophical education. This is the reason why they often confuse what they know with the actual matter in question, which, according to Popper and against Wittgenstein's opinion, constitutes the object of philosophy.

We can continue moving up the ontological chain and go from biology to psychology, or to social and historical phenomena, but we would have to repeat what we have said previously. Above the genes are the organisms, as Gould used to say. The human organism is very complex. Despite advances in neurosciences, not all psychological facts can be reduced to chemical phenomena. The explanation of neurotransmitters can partially explain some illnesses, without, to tell the truth, much success, since, for example, only 50% of depressions are cured with antidepressants, the same percentage as those cured with psychotherapy.[32] For this argument to be complete, however, it would have to explain not only "pathological" phenomena, but also "normal" ones. It would have to explain, for example, the neuroscience of marriage, patriotism, or even the minds and creativity of scientists.

I believe there would be no sense in reducing sociological and historical phenomena to mere ecologic levels, and consequently biological or physical chemical, despite efforts such as those of David Christian.[33] Without denying the value of these efforts, it seems clear from Christian's book that his "Big History" is simply a type of narration, an ecologic narration that leaves out the greater part of the dimensions of history itself, precisely because they cannot be reduced to the language of ecology.

The need for a reduction to simpler ontological or epistemological levels may be an acceptable *desideratum*, but runs the risk of becoming a sort of program of epistemological or political domination, which serves to defend the omniscience and omnipotence of what is known as science, and which, being linked at present with technological and industrial networks, could become a structure of control that is truly uncontrollable by the citizens.

In the year 2000 -- what a peculiar year -- Peter Ward and Donald Brownlee published a book that was not well received by the "scientific community."[34] These authors maintain that the Earth is a very strange planet, life is difficult to explain, and the origin of life and the present state of our species is more the result of an accumulation of chance events than of the development of a complex system of "scientific laws." Leaving aside the analysis of the facts, which in the case of Ward and Brownlee seem very convincing, what we wish to say here is that the debate centered strictly on a problem of philosophy of history and of metaphysics: the problem of necessity and contingency. These two authors advocated contingency. For them, the "standard cosmology," especially the part referring solely to the Earth, would not be the culmination of a series of processes that flow harmoniously into the present, but an accumulation of chance events, which in the final analysis would turn out to be unexplainable. This was a direct attack on the "system of science" -- which they did not wish to express thus. They were accused of defending religion and providence, although they were not doing so. Ever since the validity of narrative constructs such as "standard cosmology" and the theory of evolution began to be questioned some years ago, the champions of science accuse whoever criticizes their constructs -- not the scientific ones but those that are metascientific or narrative -- of defending creationism or the Bible, despite the fact that those critics very often are not doing so. This is because the defenders of "science" -- which does not include all scientists -- are conscious of the fact that they are taking on a role that once belonged to religion: the production of absolute truth, the final say.

Science can be revised, but not religion, they say. This is partly true. Religious texts may be interpreted within certain margins, if one accepts only the general structure of the dogma. Science can also be revised and interpreted, but also requires the acceptance of a basic dogma: its omnipotence and conclusive nature, having the final say. Apologists of science, whether scientists or philosophers of science with a bad conscience about being philosophers and begging a place under the sun within the scientific community, still want to make us believe in the PSR -- contingency's greatest enemy. The problem is that if we want to be coherent, as Pruss points out, for us to believe in that principle we must also believe in the existence of a thinking being who makes that which is thought and said coincide with what is unfortunately contingent and arbitrary.[35] That being has been called God, whether this be the God of Saint Thomas Aquinas, or the *Geist* of Hegel's *Science of Logic*, where the German philosopher -- wanting to know everything -- concluded by stating that within the development of his book one could find the same thoughts God had before the creation of the world.

Someone once asked Saint Augustine what God had thought about before creating the world. Saint Augustine answered, "the punishment he would deal out to anyone who asked such a question." We are not going to vindicate Saint Agustin or the value of theology (God forbid!), as the defenders of that argument would like to believe, but simply restore an interest in contingency, because it is in the best interest of our liberty.

And thus we conclude with an English nursery rhyme that goes something like this:

> For want of a nail the shoe was lost.
>
> For want of a shoe the horse was lost.
>
> For want of a horse the rider was lost.
>
> For want of a rider the battle was lost.
>
> For want of a battle the kingdom was lost.
>
> And all for the want of a horseshoe nail.

José Carlos Bermejo-Barrera

University of Santiago de Compostela

Spain